\documentclass[preprintnumbers,amsmath,amssymb]{revtex4}
\newcommand{\tr}{{\mathrm{tr}}}
\usepackage{graphicx}
\begin{document}

\title{JENSEN SHANNON DIVERGENCE AS A MEASURE OF THE DEGREE OF ENTANGLEMENT
%\footnote{For the
%title, try not to use more than 3 lines.
%Typeset the title in 10~pt Times roman, uppercase and boldface.}
}

\author{A.P. Majtey, A. Borras, M. Casas %\footnote{Typeset names in
%8~pt roman, uppercase. Use the footnote to indicate the
%present or permanent address of the author.}
}

\address{Departament de Fisica and IFISC-CSIC, Universitat de les Illes Balears \\
Palma de Mallorca, 07122, Spain\\
%\footnote{State completely without
%abbreviations, the affiliation and mailing address, including
%country. Typeset in
%8~pt Times italic.}\\
ana.majtey@uib.es, toni.borras@uib.es, montse.casas@uib.es}
\author{P.W. Lamberti}
\address{Facultad de Matem\'atica, Astronom\'{\i}a y F\'{\i}sica, Universidad Nacional de C\'ordoba and CONICET\\
C\'ordoba, 5000, Argentina\\
lamberti@famaf.unc.edu.ar}
\author{A. Plastino}
\address{IFLP-Facultad de Ciencias Exactas, Universidad Nacional de La Plata and CONICET\\
La Plata, 1900, Argentina\\
plastino@fisica.unlp.edu.ar}

\begin{abstract}
The notion of distance in Hilbert space is relevant in many
scenarios. In particular, ``distances" between quantum states play
a central role in quantum information theory. An appropriate
measure of  distance is the quantum Jensen Shannon divergence
(QJSD) between quantum states. Here we study this distance as a
geometrical measure of entanglement and  apply it to different
families of  states.
\end{abstract}
 \maketitle

%\keywords{Hilbert space distances; Jensen Shannon divergence;
%entanglement measures.}

\section{Introduction}
Discerning possible candidates for measuring distances between
quantum states is a subject of perennial interest. Many of these
measures  were first defined as distances between probability
distributions and subsequently employed as distance-measures in
Hilbert space. Let ${\cal H}$ be the Hilbert space associated with
a quantum system and let ${\cal S}$ be the set of all states, i.e.
the set of self-adjoint, (semi)positive and trace-one operators. A
frequently employed notion of  distance between quantum states is
the relative entropy, which is a natural extension to the realm of
quantum mechanics of the Kullback-Leibler divergence. This
quantity, however, is not useful for ascertaining the degree of
purification of an arbitrary state with respect to a pure
reference state. The relative entropy of an operator $\rho$, with
respect to an operator $\sigma$, both belonging to $\cal{S}$, is
given by
\begin{equation}
S(\rho\|\sigma)=\tr[\rho(\log\rho-\log\sigma)],\label{relative-entropy}
\end{equation}
where $\log$ stands for logarithm in base two. $S(\rho\|\sigma)$
is nonnegative and vanishes if and only if $\rho = \sigma$, being
nonsymmetric and unbounded. A particular and important requirement
indicates that the relative entropy is well defined only when the
support of $\sigma$ is equal to or larger than that of $\rho$.
Otherwise, it is defined to be $+\infty$ \cite{Lindblad} (the
support of an operator is the subspace spanned by its eigenvectors
with non-zero eigenvalues). To overcome this restriction we have
introduced a distance between elements of ${\cal S}$ that shares
with the relative entropy several of their main properties but
that is always well defined and bounded \cite{Majtey}. This
distance is the quantum Jensen-Shannon divergence (QJSD), which is
a quantum mechanical extension of the Jensen-Shannon divergence
(JSD) introduced by Rao \cite{Rao} and Lin \cite{Lin} as a
distance between probability distribution (for a detailed analysis
of the properties of the JSD, see reference \cite{Majtey0}). Here
we wish to investigate the ability of the QJSD to serve  as a
measure of the degree of entanglement.

The structure of the paper is as follows. In the next Section we
review the basic properties of the QJSD. In Section 3 we study its
properties as an entanglement measure and we apply it to quantify
the degree of entanglement of different families of two-qubit
mixed states. Finally, some conclusions are drawn in Section 4.

\section{The quantum Jensen-Shannon divergence}

We define the QJSD in the fashion \cite{Majtey}
\begin{equation}
JS(\rho\|\sigma) =
\frac{1}{2}\left[S\left(\rho\|\frac{\rho+\sigma}{2}\right)+
S\left(\sigma\|\frac{\rho+\sigma}{2}\right)\right],
\label{relative-entropy}
\end{equation}
which can be also recast in terms of the von Neumann entropy
$H_N(\rho) = - \tr(\rho \log \rho)$ as follows
\begin{equation}
JS(\rho\|\sigma) = H_N\left(\frac{\rho +
\sigma}{2}\right)-\frac{1}{2} H_N(\rho) - \frac{1}{2} H_N(\sigma).
\end{equation}
This quantity has a lot to speak for, being positive, null iff
$\rho = \sigma$, symmetric, bounded, and always well defined. In
fact, the restriction imposed on the supports of $\rho$ and
$\sigma$ for the relative entropy (\ref{relative-entropy}) is
lifted for the QJSD, that possesses  all the adequate properties
of a proper  distance between states in a Hilbert space. As stated
above, in this work we attempt to study the QJSD as an
entanglement measure which justifies listing  the main QJSD
properties. Most of these properties are discussed and proved in
Ref.~\cite{Majtey}. The list reads

\begin{itemize}

\item[(i)] $JS(\rho\|\sigma)\geq 0$ with the equality iff $\rho=\sigma$
\item[(ii)] Unitary operations left JS invariant, i.e., $JS(U\rho U^\dag\|U \sigma U^\dag) =
JS(\rho\|\sigma)$.
\item[(iii)] $JS(\tr_p\rho \|\tr_p\sigma) \leq JS(\rho \| \sigma)$ where $\tr_p$ is the partial
trace.
\item[(iv)] JS is jointly convex $JS(\sum_i \alpha_i \rho^{(i)}\|\sum_i \alpha_i \sigma^{(i)}) \leq
\sum_i \alpha_i JS(\rho^{(i)}\|\sigma^{(i)})$, where the
$\alpha_i$ are positive real numbers such that $\sum_i \alpha_i
=1$.
\item[(v)] $JS(\Phi\rho\|\Phi\sigma)\leq JS(\rho\|\sigma)$ for all
positive mappings $\Phi$.
\item[(vi)] For any set of orthogonal projectors $P_i$, such that
$P_iP_j=\delta_{ij}P_i$, $JS(\sum P_i\rho P_i\|\sum P_i\sigma
P_i)=\sum JS(P_i\rho P_i\| P_i\sigma P_i)$
\item[(vii)] $JS(\rho\otimes P_{\alpha}\|\sigma\otimes
P_{\alpha})=JS(\rho\|\sigma)$ where $P_{\alpha}$ is any projector.
\end{itemize}
The two last properties, which have not been discussed before, are
verified by the relative entropy \cite{Lindblad,Vedral98}, and
inherited by the QJSD.

In a recent work \cite{Triangular} the metric character of the
QJSD has been discussed. There it was formally proved for pure
states, and checked numerically for mixed ones by performing Monte
Carlo simulations. We can thus assert that the square root of the
QJSD verifies the triangle inequality, as it does the square root
of the JSD.

\section{The entanglement measure}

Entanglement constitutes a physical resource that lies at the
heart of quantum information processes \cite{Nielsen,LPS98,BEZ00}.
Quantum teleportation, superdense coding, and quantum computation,
are some of the most representative examples. Let us recall that a
state of a composite quantum system is called entangled if it can
not be expressed as a convex sum of factorizable pure states.
Otherwise, the state is called separable.

Nowadays a variety of measures are used to quantify the degree of
entanglement. These include the entanglement of distillation, the
relative entropy of entanglement, etc. The canonical measure of
entanglement in a bipartite pure systems is the so-called
entanglement of formation, which is a strictly monotonic function
of the squared concurrence. For simplicity the entanglement of
formation is frequently used as {\it the} measure of entanglement.
However, for mixed state there exist several available measures.

Before starting with the entanglement-characterization using the
QJSD we study the structure of Hilbert space (HS) according to
this distance measure. To do that we evaluate $d_{JS}=\sqrt{JS}$
between a given random generated state and the maximally mixed
(MM) state $\rho_{MM}=I/N$, with $N$ the HS-dimension. In Fig. 1
we plot the probability distribution for finding an arbitrary
state $\rho$ at a given distance from the $\rho_{MM}$. We find, as
 expected, that the mean value of such distance increases for
higher Hilbert space-dimensions.

Let us now enumerate the properties that any adequate measure of
entanglement should satisfy \cite{Bengtsson}:

\begin{itemize}
 \item[(i)] Discrimination: ${\cal{E}}(\rho)=0 \;\; iff\;\; \rho$ is separable.
 \item[(ii)] Monotonicity: the measure does not increase under local
 general measurements and classical communication, for every
 completely positive map $\Phi$,
 ${\cal{E}}(\Phi\rho)\leq{\cal{E}}(\rho)$
 \item[(iii)] Convexity: ${\cal{E}}(x\rho + (1-x)\sigma)\leq
 x{\cal{E}}(\rho)+(1-x){\cal{E}}(\sigma)$, with $x \in [0,1]$.
\end{itemize}

As it was already stated, the main purpose of the present
communication is to investigate the QJSD as a {\it geometrical}
measure of entanglement. Following Vedral \textit{et al.},
\cite{Vedral97} we define an entanglement measure $\cal{E}(\rho)$
as the minimum QJSD from the state $\rho$ to the set ${\cal D}$ of
the disentangled states.
\begin{equation}
{\cal{E}}_{JS}(\rho)=\min_{\sigma\in {\cal
D}}JS(\rho,\sigma).\label{js-entanglement}
\end{equation}
The QJSD properties enunciated in the preceding section ensure
that  (\ref{js-entanglement}) fulfills the conditions for an
adequate entanglement measure. For convenience's sake, we
normalize the entanglement measure by a trivial re-scaling in
order to adequately compare different quantities of interest,
i.e.,
${\cal{E}}(\rho)={\cal{E}}(\rho)/{\cal{E}}(|\psi^-\rangle\langle\psi^-|).$

It is notheworthy to stress that expression
(\ref{js-entanglement}) has already been investigated for
different distances, other than the JS-one. For example, Vedral
and coworkers used to this effect the relative entropy
(${\cal{E}}_{RE}(\rho)$) and the Bures metric
(${\cal{E}}_B(\rho)$) \cite{Vedral97,Vedral98}. Also the
Trace-distance and the Hilbert-Schmidt metric have received
consideration for the purpose \cite{Eisert03,Witte99}. We shall
below use as a reference the quantity ${\cal{E}}_B(\rho)$.
\begin{figure}[h]
\includegraphics[scale=0.9,angle=0]{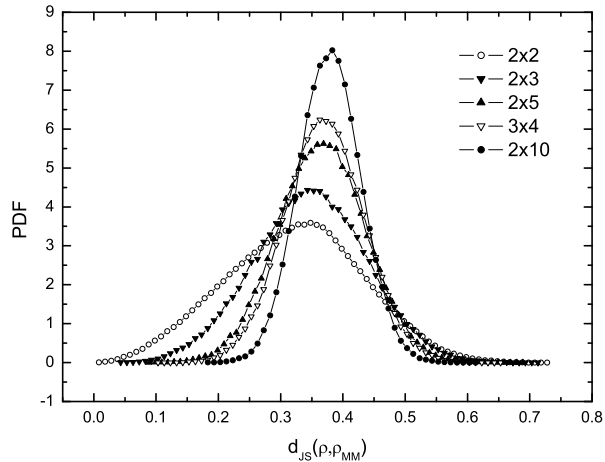}
\caption{Probability distribution for finding a quantum state at a
given distance from de maximally mixed one (for different Hilbert
space dimensions).}
\end{figure}
\begin{figure}
\includegraphics[scale=0.9,angle=0]{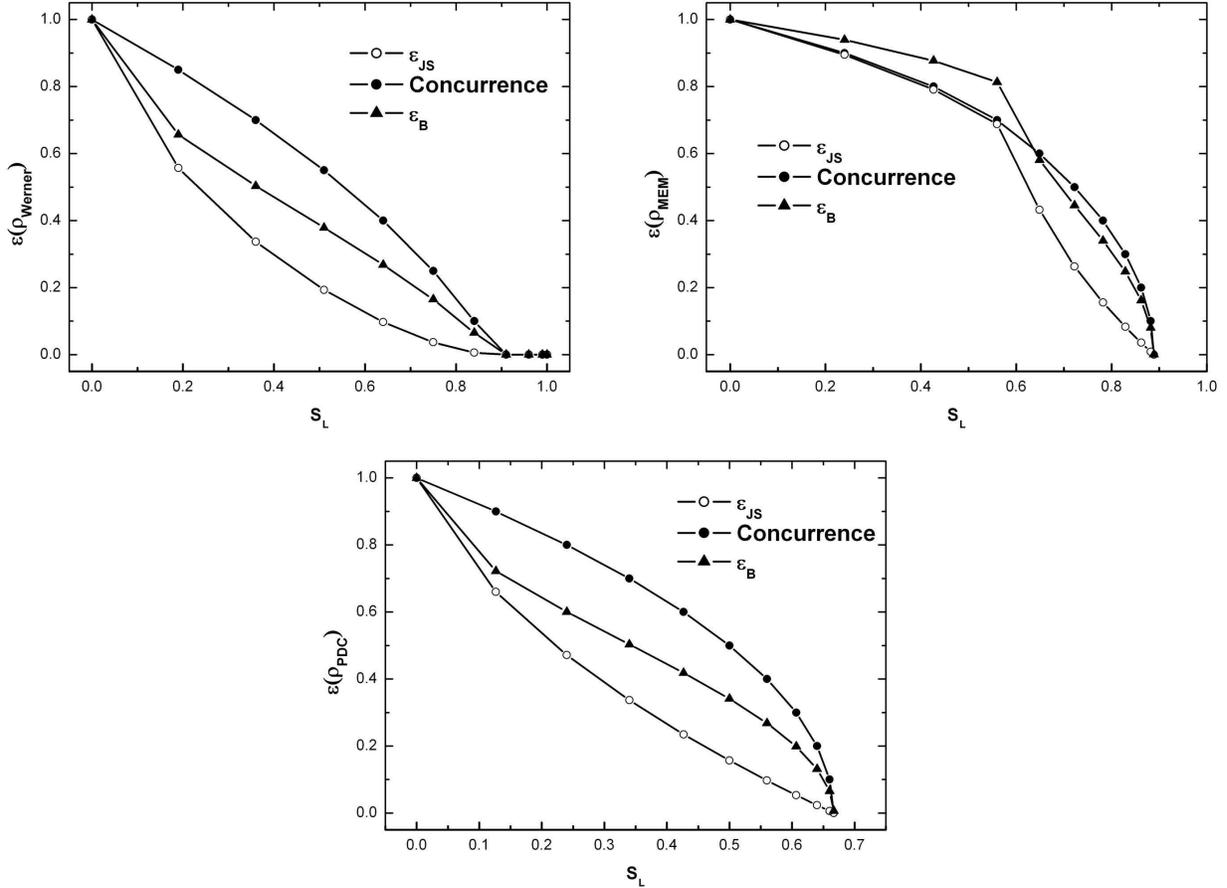}
\caption{Comparison between the concurrence, the JS- measure, and
the measure  ${\cal{E}}_{B}$ induced by the Bures metric. The task
is accomplished for (a) Werner, (b) MEM, and (c) PDC states.}
\end{figure}
In order to investigate the behavior of the QJSD as an
entanglement measure we consider now three well-known families of
two-qubits mixed states, namely, \begin{itemize} \item Werner ones
\cite{Werner}, \item maximally entangled mixed ones \cite{MEM}
(MEM), and \item parametric down-conversion \cite{PDC} (PDC)
states.
\end{itemize} All of these are diagonal in the Bell basis. We perform an
optimization procedure so as to find the  separable state that
lies in the closest proximity to the state of interest $\rho$.
Such task is performed by following a simulated annealing
minimization procedure, starting from the maximally mixed state
$\rho_{MM}$. Appropriately perturbing $\rho_{MM}$ we look for the
minimal QJSD between $\rho$ and some state in $\cal{D}$.

In Fig.2 we depict, versus the linear entropy $S_L$ (the degree of
impurity), i) the concurrence and the two normalized quantities:
ii) our ${\cal{E}}_{JS}$ and iii) Vedral's ${\cal{E}}_{B}$
measure. This is done for the three families enumerated above.

Notice that the concurrence and  ${\cal{E}}_{B}$ are greater than
${\cal{E}}_{JS}$, except for the ``extreme" cases of maximally
entangled or disentangled Bell states in two of the sub-figures.
In the MEMs-instance this is not entirely so. Crossings are
detected between ${\cal{E}}_{B}$ and the concurrence ${\cal{C}}$,
although the ${\cal{E}}_{JS}-$curve lies always below the other
two for all the three families, except for the extreme (trivial)
situations. The behavior here described mimics that of the
relative entropy vs. $S_L$. In that instance,  the relative
entropy-induced entanglement is strictly smaller than that of
formation, as analytically demonstrated in \cite{Vedral98}.

Our crucial point is the fact that the measure induced by the QJSD
is easier to compute than that induced by the Bures metric. This
is so because of the square root in the definition of Bures'
metric, which forces cumbersome matrix-diagonalization and basis'
changes in order to compute ${\cal{E}}_{B}$. Instead, so as to
calculate the QJSD one i) only needs to solve the
eingenvalue-equation for three matrices (eigenvectors are not
necessary) and ii) the optimization procedure is much more
efficient than in the Bures situation.

\section{Conclusions}

We have here advanced a new entanglement-measure ${\cal{E}}_{JS}$,
based on the Jensen-Shannon distance. For three important families
of states this new quantity behaves in rather similar fashion as
the established entanglement-measure ${\cal{E}}_{B}$ of Vedral
\textit{et al.}, with an important difference: the former is much
more easy to compute than the later.

\section*{Acknowledgments}

This work was partially supported by the MEC grant FIS2005-02796
(Spain) and FEDER (EU) and by CONICET (Argentine Agency). AB and
APM acknowledge support from MEC through FPU grant AP-2004-2962
and contract SB-2006-0165. PWL wants to thank SECyT-UNC
(Argentina) for financial support. We thank J. Batlle and Prof. A.
R. Plastino for using his Hilbert-Monte Carlo numerical program.

\end{document}